\documentclass[%
 reprint,
superscriptaddress,
nofootinbib,
showkeys,
 amsmath,
 amssymb,
 aps,
prl,
nobibnotes,
floatfix,
longbibliography,
]{revtex4-2}

\usepackage{graphicx, subfigure}
\usepackage{dcolumn}
\usepackage{bm}
\usepackage{color}
\usepackage{booktabs}
\usepackage{mathrsfs}[euscript]
\usepackage{enumitem}
\usepackage{tikz}
\usepackage{array}
\usepackage{boldline,multirow, hhline}
\usepackage{eucal}
\usepackage{notes2bib}
\usepackage{placeins}
\usepackage[hidelinks]{hyperref}
\usepackage{filecontents}
\niibsetup{writekey=false,name={}}

\hypersetup{
  colorlinks  = true, 
  urlcolor    = blue, 
  linkcolor   = blue, 
  citecolor   = red ,  
  filecolor    = blue
}
\newcolumntype{M}[1]{>{\centering\arraybackslash}m{#1}}
\newcolumntype{N}{@{}m{0pt}@{}}

\niibsetup{writekey=false,name={}}
\begin{document}

\preprint{AIP/123-QED}

\title[Kulkarni et al.]{Coalescence and spreading of drops on liquid pools}

\author{Varun Kulkarni}
\altaffiliation[Present address: ]{School of Engineering and Applied Sciences, Wyss Institute for Biologically Inspired Engineering, Harvard University, Cambridge, MA 02138}
\email[\\ Corresponding author: ]{varun14kul@gmail.com}
\author{Venkata Yashasvi Lolla}
\altaffiliation{Present address:  Department of Mechanical Engineering, Virginia Polytechnic Institute and State University, Blacksburg, VA 24060}
\altaffiliation{\\Equal contribution as second author}
\author{Suhas Tamvada}
\altaffiliation[Present address: ]{Department of Mechanical and Aerospace Engineering, University of Florida, Gainesville, FL 32611}
\altaffiliation{\\ Equal contribution as second author}
\author{Nikhil Shirdade}
\altaffiliation[Present address: ]{Department of Mechanical Engineering, Pennsylvania State University, State College, PA 16802}
\author{Sushant Anand}
\email[Corresponding author: ]{sushant@uic.edu}
\affiliation{Department of Mechanical and Industrial Engineering \\ University of Illinois, Chicago, IL 60607}


\begin{abstract}
Oil spills have posed a serious threat to our marine and ecological environment in recent times. Containment of spills proliferating via small drops merging with oceans/seas is especially difficult since their mitigation is closely linked to the coalescence dependent spreading. This inter-connectivity and its dependence on the physical properties of the drop has not been explored until now. Furthermore, pinch-off behavior and scaling laws for such three-phase systems have not been reported. To this end, we investigate the problem of gentle deposition of a single drop of oil on a pool of water, representative of an oil spill scenario. Methodical study of 11 different \textit{n}-alkanes, polymers and hydrocarbons with varying viscosity and initial spreading coefficients is conducted. Regime map, scaling laws for deformation features and spreading behavior are established. The existence of a previously undocumented regime of \textit{delayed} coalescence is revealed. It is seen that the inertia-visco-capillary (I-V-C) scale collapses all experimental drop deformation data on a single line while the early stage spreading is found to be either oscillatory or asymptotically reaching a constant value, depending on the viscosity of the oil drop unlike the well reported monotonic, power law late-time spreading behavior. These findings are equally applicable to applications like emulsions and enhanced oil recovery.
\end{abstract}

\pacs{Valid PACS appear here}
\maketitle

Oil-water interactions play a crucial role in numerous industries such as, pharmaceutical, cosmetic, petrochemical, agrochemical, and food processing - encompassing applications involving emulsions \cite{Zhu2020, Utada2005, Guha2017} for drug delivery \cite{Lawrence2000, Dittrich2006}, ointments \cite{Thivilliers2006, Bolleddula2010}, paint sprays \cite{Yang2018b},  salad dressings \cite{Franco1995}, pesticides \cite{Feng2018}, microfluidic multiphase reactors \cite{Song2006, Teh2008}, material synthesis \cite{Imhof1997, Marre2010}, thermofluids \cite{Smith2013}, CO$_2$ sequestration and enhanced oil recovery \cite{Lemahieu2019}. These interactions are also central to many problems we face today, a salient example of which are the oil spills due to leakages from oil tankers aboard maritime vessels. In addition to the large volumes spilled during such an event, a substantial amount also enters water bodies in the form of impacting oil droplets (emerging from fragmentation of leaking jets of oil). While the behavior of spilled oil over vast regions over oceans and seas has been studied extensively \cite{Gong2014, Fay1971,Hoult1972}, several facets of oil drops impacting water have remained unexplored despite drop impact being one of the most active areas of research \cite{Yarin2006, Rein1993}. 

Studies on drop impact have revealed it can be characterized in terms of its several constituent elements, namely, the physical properties of the liquid drop \cite{Yarin2006}, impact velocity \cite{Clanet2004}, impact angle \cite{Almohammadi2017}, ambient pressure \cite{Xu2005}, and the chemical/physical attributes of the surfaces which could be smooth/rough \cite{Clanet2004}, soft/elastic \cite{Howland2016}, or a deep pool \cite{Rein1993} or thin film of liquid \cite{Huang2008}. Among these many problems, our interest lies specifically in studying how an oil droplet gently impacting on water transforms into an oil film and how its initial spreading behavior can be understood - aspects that are critical to oil spills. The model problem we study represents a coalescence event involving two entities - a finite sized spherical droplet and a pool of liquid representing a planar surface. Also, simultaneous to the coalescence is the spreading of the oil drop over the pool of water underneath which encompasses an initial time and late time behavior occurring at differing length and time scales. While studies to date \citep{Aryafar2006a, Blanchette2006} have considered merger of droplet of one fluid with itself (in bulk form), our model problem involves an advancing three-phase contact line comprising an oil-water-air interface which influences the initial time spreading behavior dominantly - a problem which has largely remained unanswered until now.

Prior work on drop coalescence with pools of liquids (in three-phase systems) dates back to the research by eminent scientists such as Osborne Reynolds \cite{Reynolds1875} and J.J. Thomson \cite{Thomson1886} in late 1800s who showed that such coalescence events proceed in stages wherein a secondary (daughter) droplet - fraction of the size of the original parent drop may also be produced and \cite{Thomson1886} produces vortex rings deep inside the pool. These initial forays were followed by the celebrated experiments of A.M. Worthington in 1908 who examined impact of a drop of milk on a pool of water which confirm these observations and are detailed in the epochal monograph, \textit{The study of splashes} \cite{Worthington1908}.  However, the exact mechanism of coalescence remained unknown until the works of Cockbain and McRoberts \cite{Cockbain1953} and  Gillespie and Rideal \cite{Gillespie1956} who were the first ones to examine the role of the intervening air film, its drainage and rupture in a two- phase system. Charles and Mason \cite{Charles1960a} then advanced these ideas coining the term \textit{partial coalescence} for  merger that occurs in stages and attributed the formation of a daughter droplet to Rayleigh instability. These investigations \cite{Cockbain1953, Gillespie1956, Charles1960a, Charles1960b} in the context of gentle impact of droplets on a pool of the same liquid (as the drop) revolved around three major axes - understanding the mechanism and forces underlying the formation of the daughter droplet and prediction of its size; determining the conditions leading to partial and complete coalescence; time for complete coalescence. However, it was not until the advent of advanced high-speed imaging in the 2000s that much of these nuances of partial coalescence came to light \cite{Aryafar2006a, Chen2006a, Blanchette2006, Blanchette2009a}. 

The key findings from these studies, all of which were for two-phase systems, was that the ratio($D_d/D_p$) of the size (diameter) of the daughter droplet ($D_d$) to the parent drop ($D_p$) and the coalescence time ($t_c$) is a function of the liquid properties given by the drop Ohnesorge number, $Oh_p$ ($=\mu/\sqrt{\rho_p \gamma D_p}$), where $\mu_{p}$, $\rho_{p}$ and $\gamma$ are the liquid viscosity, density and surface tension respectively. Chen et al. \cite{Chen2006a,Chen2006b,Aryafar2006a} included the effects of gravity by introducing the non-dimensional Bond number,  \textit{Bo} ($=\rho_p g {D_p}^2/\gamma$) and identified three distinct boundaries for inertia, visco-capillary and gravitational regimes. Furthermore, they presented phase diagram in terms of $Bo-Oh$ demarcating regions of partial and complete coalescence. Similar analysis was presented by Gilet et al. \cite{Gilet2007a}. Finally, Blanchette and Biogini \cite{Blanchette2006, Blanchette2009a} questioned the validity of Rayleigh instability \cite{Charles1960a, Charles1960b} as being responsible for droplet generation and instead posited an explanation based on the development of capillary waves which vertically stretch the drop by focusing energy on its summit leading to the pinch off of a daughter droplet. Compared to two-phase coalescence problems herein we elucidate the role of capillary waves in sculpting the main features of drop deformation and pinch off in three-phase, oil-water-air system through previously unknown phase diagram and scaling relations dependent only on the liquid properties. 

As mentioned before, the coalescence event occurs concurrently with the spreading of the oil drop on the water surface. Much of
the research so far has focused on coalescence without describing its consequences on spreading behavior. The spreading characteristics such as maximum spread area and spreading rate are vital to applications like oil recovery \cite{Lemahieu2019}, drug encapsulation and delivery \cite{Dittrich2006} besides oil spill remediation \cite{Broje2007,Riehm2017}. 

Central to the spreading of oil on a liquid surface is the concept of an equivalent force at the three-phase contact line due to the interfacial tensions, often described by the initial spreading coefficient, $S = \sigma_{wa} - \sigma_{oa} - \sigma_{ow}$ \cite{Craster2006, DeGennes2013}, where $\sigma$ is the surface/interfacial tension and the subscripts \textit{wa}, \textit{oa}, and \textit{ow} denote the water-air, oil-air, and oil-water interfaces respectively. A positive value of $S$ denotes an affinity of the liquid to spread \cite{DeGennes2013} on the substrate while a negative value implies a tendency of the mass of oil to form a liquid lens \cite{Langmuir1933}. For spreading oils ($S > 0$) a two-dimensional monolayer \cite{DeGennes2013}, known as precursor film precedes the macroscopic liquid front. Studies by Huh et al. \cite{Huh1975} and Joos et al. \cite{Joos1977,Joos1985} showed that its spreading is governed by a competition between the interfacial tension and viscous forces which result in a spreading radius, $r\left(t\right)$ variation with time as, $r\left(t\right) \sim t^{3/4}$. Some spreading liquids may also exist in partial-wetting regime wherein they may show a pseudo-lens along with a thin film because of long-range forces \cite{Ragil1998}. Regardless, for spreading oils with a film thickness of $\mathcal{O}$($10^{-6}$) \cite{Fay1971,Hoult1972} the radial spread has been shown to follow power law behavior, $r\left(t\right) \sim t^n$ \cite{Huppert1982,Maderich2012}. Fay and Hoult \cite{Fay1971} provide a comprehensive summary of values for \textit{n}, which are found to lie between 3/8 and 2/3 depending on which forces are dominant - inertial, viscous or surface tension and also whether the film is considered axisymmetric or one dimensional.  

A common feature of the above-mentioned previous studies is that they mainly investigated the late time behavior which is relatively unaffected by early time occurrences where the coupling of the coalescence phenomenon with spreading is evident. The early time spreading can affect the overall extent to which a film can spread and has so far remained unexplored. 

Moreover, understanding them might inform development of oil spill mitigation measures. Consequently, in the present study we aim to unravel this initial time macroscopic behavior of oil drops spreading on water with the objective to contrast it with long time behavior and highlight the differences.

In summary, we identify the following key objectives for our work: establish scaling laws for various features of the drop as it deforms during coalescence, including the size of the daughter droplet and capture the initial spreading dynamics which we report to be markedly different from the long-time behavior.

\section{\label{experimental}Materials and Methods}
\vspace{-5pt}
\subsection{Chemicals used}
\vspace{-5pt}
To consider the influence of wide ranging viscous and interfacial properties a variety of oils were chosen. The alkanes - Hexadecane (ReagentPlus $99\%$), Kerosene (Reagent Grade), Pentane (Reagent, $98\%$), Decane (Reagent, $\geq 95\%$), Dodecane (ReagentPlus, $\geq 99\%$), Tetradecane $\left(\geq 99\%\right)$, and Cyclohexane (ACS Reagent, $\geq 99\%$) were chosen for their relative abundance in crude oil \cite{Drelich1999} and  purchased from Sigma Aldrich \textregistered (St. Louis, Missouri, USA ). Silicone oils (SO) of varying viscosity (0.65 cSt, 1.5 cSt, 10 cSt,100 cSt) were purchased from Gelest Inc.\textregistered (Morrisville, Pennsylvania, USA) to study viscosity effects. Table \ref{tab:Table1} illustrates the properties of the liquids used including kinematic viscosity, $\nu$ of the oils, interfacial tensions ${\sigma}_{aw}$, ${\sigma}_{oa}$, ${\sigma}_{ow}$, and the spreading coefficient \textit{S}, where, subscripts \textit{ow}, \textit{oa}, and \textit{aw} denote the oil-water, oil-air, and air-water interfaces respectively. 

Oils may exist in different wetting states depending on the value of $S$ and Hamaker constant ($A$). In particular, note that pentane exists in a partial-wetting state on water \cite{Ragil1998} (see details in Section S6, SI). The density of air and water are taken to be 1.29 $kg/m^3$ and 998 $kg/m^3$ while their dynamic viscosities are considered as $18.1\;\mu Pa-s$ and $1\;mPa-s$ \cite{Davis2010} respectively. For the oils, the density is estimated by dividing a known mass of the liquid and dividing it by its volume (Table \ref{tab:Table1}). The measurements thus obtained are within an uncertainty of 5\%.

The coalescence and spreading process studied in this work occurs in $\leq 1s$ (see Fig. \ref{fig:Figure2}) which is insufficient for the impurities to be adsorbed at the water-oil interface and influence the spreading behavior \cite{Burnett2004}. The physical properties of the liquids thus may be equivalently expressed in terms of non-dimensional quantities such as the Ohnesorge number of the parent drop $\left(Oh_p\right)$, Bond number $\left(Bo_p\right)$, viscosity ratio $\left(_a\mu_r\right)$ and $\left(_b\mu_r\right)$ and density ratio $\left(_a\rho_r\right)$ and $\left(_a\rho_r\right)$ (see Section S2, SI, Table S1 for details on non-dimensional quantities and Section \ref{sec:danalysis} for definition of these quantities).

Deionized water was used as the bulk liquid, and was held in a quartz chamber of dimensions $30\times30\times30$ $mm$ which was treated with Trichloro [1H, 1H, 2H, 2H - Peflorooctyl silane purchased from Sigma Aldrich \textregistered to prevent the formation of a concave or convex meniscus along the glass walls by maintaining a 90$^\textrm{o}$ contact angle(hydrophobic). Vapor salinization of the glass chamber ensured any image distortions due to the presence of a meniscus during side view imaging were avoided.
\begin{table}[htbp!]
\begin{center}
{ 
\vspace{-5pt}
\caption{Physical properties of the oils used in the study. $\rho$ is density, $\nu$ is kinematic viscosity, $\sigma$ is interfacial tension (different from $\gamma$ which is the equivalent interfacial tension defined in section \ref{sec:danalysis}), \textit{S} is the initial spreading coefficient, ($= \sigma_{wa} - \sigma_{oa} - \sigma_{ow}$) where $\sigma_{aw} =$ 72 $mN/m$. No reference is provided for values of density which are measured in the laboratory and some values for silicone oils which are directly available in the Gelest Inc.\textregistered (Morrisville, Pennsylvania, USA) brochure.}\label{tab:Table1}
\resizebox{\columnwidth}{!}
{\color{black}\begin{tabular}{l*6c}
\toprule
  \multirow{2}{*}{Liquid} & \hspace{0.3em} ${\rho}$ & \hspace{0.3em} ${\nu}$ & \hspace{0.3em} ${\sigma}_{oa}$ & \hspace{0.3em} ${\sigma}_{ow}$ & \hspace{0.3em} \textit{S} \\
    &\hspace{0.3em} ($kg/m^3$) &\hspace{0.3em} ($mm^2/s$) & \hspace{0.3em}$(mN/m)$ & \hspace{0.3em}$(mN/m)$ & \hspace{0.3em}$(mN/m)$\\
\midrule
     Silicone Oil  & \hspace{0.3em} 761 & \hspace{0.3em} 0.65  & \hspace{0.3em} 15.9 & \hspace{0.3em} 38.7 \cite{Binks2002} & \hspace{0.3em} 17.4\\
     \\[-1em]
     Silicone Oil & \hspace{0.3em} 853 & \hspace{0.3em} 1.5  & \hspace{0.3em} 17.8  & \hspace{0.3em} 42.5 \cite{Kanellopoulos1971} & \hspace{0.3em} 11.7\\
     \\ [-1em]
     Kerosene & \hspace{0.3em} 810 & \hspace{0.3em} 2.39 \cite{Abulencia2009}  & \hspace{0.3em} 28.0 \cite{Abulencia2009}  & \hspace{0.3em} 33.0 \cite{Wan2019}  & \hspace{0.3em} 11.0\\
     \\ [-1em]
     Silicone Oil & \hspace{0.3em} 935 & \hspace{0.3em} 10  & \hspace{0.3em} 20.1 	& \hspace{0.3em} 43.0 \cite{Wehking2014} & \hspace{0.3em} 8.9\\ 
     \\[-1em]
     Silicone Oil & \hspace{0.3em} 966 & \hspace{0.3em} 100 	& \hspace{0.3em} 20.9  & \hspace{0.3em} 43.1 \cite{Wehking2014} & \hspace{0.3em} 8.0\\
     \\ [-1em]
     Pentane & \hspace{0.3em} 626 & \hspace{0.3em} 0.5 \cite{Yaws2003} & \hspace{0.3em} 15.5 \cite{Demond1993}  & \hspace{0.3em} 49.0 \cite{Demond1993} & \hspace{0.3em} 7.5 \\ 
     \\ [-1em]
     Decane & \hspace{0.3em} 726 & \hspace{0.3em} 1.27 \cite{Yaws2003} & \hspace{0.3em} 23.83 \cite{Demond1993}  & \hspace{0.3em} 52.0 \cite{Demond1993}  & \hspace{0.3em} -3.83 \\ 
     \\ [-1em]
     Cyclohexane & \hspace{0.3em} 773 & \hspace{0.3em} 1.15 \cite{Yaws2003}  & \hspace{0.3em} 26.56 \cite{Demond1993}  & \hspace{0.3em} 50.0 \cite{Demond1993}  & \hspace{0.3em} -4.56 \\ 
     Dodecane  & \hspace{0.3em} 746 & \hspace{0.3em} 2.01 \cite{Yaws2003} & \hspace{0.3em} 24.91 \cite{Demond1993}  & \hspace{0.3em} 52.8 \cite{Demond1993} & \hspace{0.3em} -5.71 \\ 
     \\ [-1em]
     Tetradecane & \hspace{0.3em} 756 & \hspace{0.3em} 3.06 \cite{Yaws2003} & \hspace{0.3em} 26.6 \cite{Demond1993}  & \hspace{0.3em} 52.2 \cite{Demond1993} & \hspace{0.3em} -6.8\\ 
     \\ [-1em]
     Hexadecane & \hspace{0.3em} 770 & \hspace{0.3em} 4.46 \cite{Yaws2003} & \hspace{0.3em}26.95 \cite{Demond1993}  & \hspace{0.3em}53.3  \cite{Demond1993}  & \hspace{0.3em} -8.25 \\
\bottomrule
\end{tabular}
}
}
\end{center} 
\vspace{-20pt}
\end{table}
\vspace{-10pt}
\subsection{Experimental Setup} \label{subsec:exp_setup}

\begin{figure} [htpb!]
\includegraphics[width=\columnwidth]{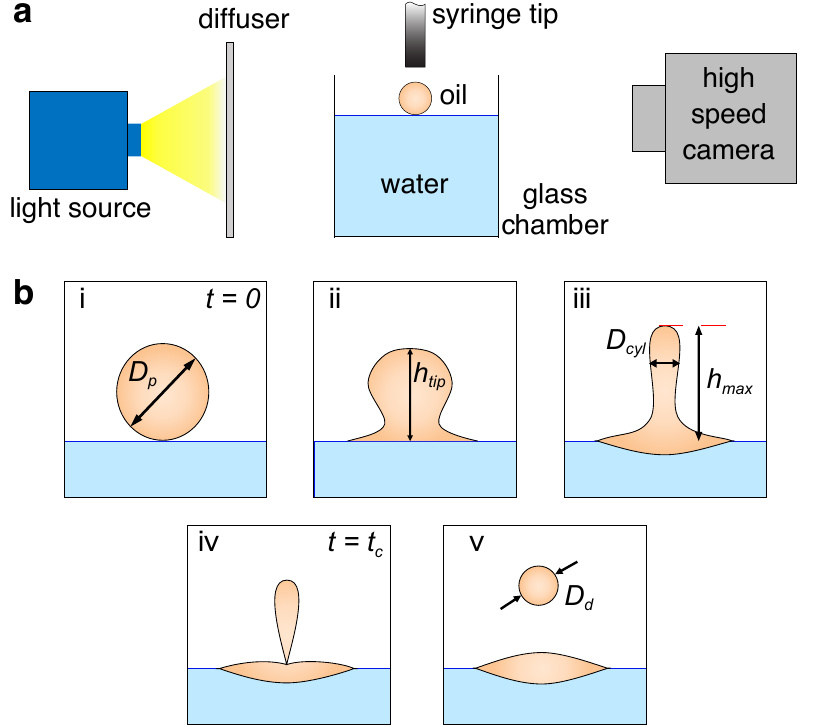}
\centering
   \caption{\label{fig:Figure1} \textbf{(a)} A schematic of the setup apparatus used to record videos from the front. Standard high-speed imaging was used to capture drop dynamics during partial coalescence. The diffuser plates are used to uniformly distribute the light from the source. \textbf{(b)} The various parameters studied during partial coalescence, namely parent drop diameter $D_p$, drop height at any instance in time $h_{tip}$, maximum height of drop $h_{max}$, coalescence time $t_c$, diameter of daughter droplet $D_d$, diameter of the upward cylinder $D_{cyl}$.}
\end{figure}

The schematic in Fig. \ref{fig:Figure1}(a) shows the components of the experimental setup used to record videos from the front and observe coalescence events. Droplet spreading was captured from the top by orienting the camera vertically towards the water surface. Needles of sizes varying from 0.6-1.6 mm ID corresponding to gauge sizes of 14, 16, 18, and 20 were attached to a glass syringe for generating drops. The syringe was mounted onto a syringe pump and operated at a flow rate of 1.67 $\mu l/s$ to avoid any vibrations from the stepper motor housed within the syringe pump and ensure consistent drop sizes for a particular needle size. The entire apparatus was placed on an optical table to eliminate mechanical vibrations. All experiments were conducted at a temperature of 25 $\pm$ 3 $^\textrm{o}$C.

The drops (referred to as \textit{parent drops} in the remaining text) were gently deposited ($<$ 5 $mm/s$) on a pool of water to minimize any inertial effects. The depth of the pool was about $25\;mm$ and it was large enough to ensure that the dynamics of the overlaying oil drop was not obscured by presence of a bottom plate \cite{Tropea1999}. The drops ($R_p = D_p/2$) generated by the nozzles were such that they are smaller than the capillary length, $l_c \left(= \sqrt{\gamma/\rho_pg}\right)$ where, $\gamma$ is equivalent interfacial tension of the drop (see further in Section S3, SI), $g$ is the gravitational acceleration and $\rho_p$ is the density of the parent drop, thus ensuring that gravity played no role in the deformation dynamics. To ensure that the water-air interface was perfectly flat so that a deposited droplet does not move towards the wall due to capillary attraction \cite{Kralchevsky1994}, the glass chamber size was made $\approx$ 10 times larger than typical drop sizes and the walls were silanized. The top/side view experiments were performed independent of each other \cite{Feng2014} since the deformation was found to be repeatable over a number of experiments with an experimental uncertainty of less than 7\%.

In order to capture the coalescence and spreading behavior, high-speed imaging using Photron \textregistered FASTCAM Mini AX camera was employed. Videos were recorded at 4000 frames per second (\textit{fps}) at a resolution of 1024 $\times$ 1024 pixels with an exposure time of 5 $\mu s$. This was deemed sufficient for capturing the entire dynamics given the temporal resolution was about 0.25 $ms$ and the typical time scale for the entire process was in excess of 10 $ms$ (see Fig. \ref{fig:Figure2} (a),(b),(c)). A lens with infinite focus (InfiniProbe \textregistered TS$-$160) with focal length ranging from infinity to 18 $mm$ and a magnification ranging from 0$-$16$\times$ was attached to the camera resulting in the arrangement providing a magnification of 1 pixel $\approx$ 10 $\mu m$. An LED (Nila-Zaila\textregistered) illumination source was used and the emanating light was diffused using multiple diffuser plates placed between the light source and the glass chamber.

The videos were recorded to capture the drop deformation dynamics from the moment the parent drop touches the pool of water to its complete coalescence with the water.The videos were analyzed using the open source IMAGE J software. Fig. \ref{fig:Figure1}(b) illustrates some of the parameters investigated during drop coalescence and includes (\textit{i}) parent drop diameter, $D_p$, (\textit{ii}) height of the collapsing drop at any instance in time, $h_{tip}$, (\textit{iii}) the maximum height of the drop, $h_{max}$, (\textit{iv}) diameter of the upward stretched cylinder (jet), $D_{cyl}$ (\textit{v}) the time taken for drop coalescence, $t_c$, and (\textit{vi}) daughter drop diameter, $D_d$. Similarly, Fig. \ref{fig:Figure5} provides details of quantities studied to understand the dynamics of the spreading process.
\section{Results} \label{sec:results}

As an oil drop gently impacts a bulk liquid (water) it levitates over the liquid surface for some moments before integrating with the bulk \cite{Kavehpour2015, Chen2006b}. The delay in coalescence is due to an intervening air film sandwiched between the oil drop and bulk liquid \cite{Kavehpour2015}. As the air film drains gradually it eventually ruptures and leads to contact between the two liquids\cite{Brown1967,Charles1960b, Charles1960a}. Concurrent to the rupture of this air film is the  initiation of capillary waves which propagate along the phase boundary of the drop and ambient medium, deforming the drop in the process \cite{Gilet2007a, Charles1960b, Charles1960a, Chen2006b, Kavehpour2015}. The strength of the capillary waves, which is a measure of the energy they possess is responsible for the observed drop deformation and coalescence. In these initial moments, the coalescence may also affect spreading of the drop over the liquid pool. The global topological variables such as total time for coalescence ($t_c$), maximum vertical stretch, ($h_{max}$) and the daughter droplet radius ($R_d$) thus strongly depend on the physical properties of the liquid drop.

To probe these aspects in detail and elucidate the key mechanisms we begin by identifying the key dimensionless parameters and dimensional scales. Thereafter, we use this information to delineate different regimes of coalescence by means of a regime map followed by developing scaling laws for the free surface features and conclude by studying their effect on the spreading kinetics.
\vspace{-10pt}
\subsection{Dimensional analysis}\label{sec:danalysis}

The capillary waves causing drop deformation during its coalescence with water (bulk) generate free surface deformations ($\Omega$) which primarily depend on the parent drop diameter ($D_p$), interfacial tension ($\gamma$), density ($\rho_b$, $\rho_p$) and viscosity ($\mu_b$, $\mu_p$) of the bulk and liquid drop, velocity of impact ($U$) and the magnitude of gravitational acceleration ($g$). In reality, we may have a dependence on a few more variables than shown here because we have a three-phase system and each of the phases - air, bulk and the ambient have their own density and viscosity. However, most studies \cite{Chen2006a, Chen2006b, Aryafar2006a} thus far have been conducted on just two-phases. Even in our case the capillary waves essentially travel along the two-phase boundary of the air and the oil drop. So, to avoid repetition we consider only density and viscosity of the drop and the bulk with the understanding that to include the effects of the ambient we would just need to change the fluid properties from bulk to the ambient. Therefore we may now consider the drop deformation features (\textit{i}-\textit{v}, section \ref{subsec:exp_setup}) to be only a function of the 8 primary variables of the problem. 
\begin{equation} \label{eqn:dimvar}
\Omega \sim f({\rho}_{p}, {\rho}_{b}, {\mu}_{p}, {\mu}_{b}, {\gamma}, {D}_{p}, g, U)
\end{equation}
Here, $\gamma$ is the equivalent interfacial tension (see Section S3, SI for details) of the drop and defined separately for spreading and non-spreading oils as ,
\begin{equation} \label{eqn:gamma}
\gamma =
\left\{
	\begin{array}{ll}
		(\sigma_{oa} + \sigma_{ow})/2  & \mbox{if}\;\;\;\;S > 0 \\
		\qquad |S| & \mbox{if}\;\;\;\;S < 0
	\end{array}
	\right.
\end{equation}
Applying Buckingham's Pi {$\left(\pi\right)$} theorem \cite{Barenblatt1996} the dimensional dependence in eqn \ref{eqn:dimvar} leads to, $8 - 3 = 5$ $\pi$ terms which is the difference of the 8 variables of our problem and the three fundamental units of mass, length and time. Each of these $\pi$ terms is a non-dimensional group and found to be, $Bo_p = \Delta \rho_p g D_p^2/\gamma$, $Oh_{p,b} = \mu_{p,b}/\sqrt{\rho_{p,b}\gamma D_p}$, $We_p= \rho_p U^2 D_p/\gamma$, $\rho_r = \rho_p/\rho_b$ and $\mu_r=\mu_p/\mu_b$, where, subscripts $p$ and $b$ refer to the parent drop and the bulk, respectively. $We_p$ for our case for a velocity, $U < 5 mm/s$ is less than 5 and considered negligible. Similarly, $Bo_p$ is insignificant ($\lessapprox 1$) for our case too (see Section S2, SI for further details). Equivalently, it means $D_p < \ell_c$. Note that, $Oh_p = Oh_b \left(\mu_r/\sqrt{\rho_r}\right)$ so it suffices to use one of the variables, $Oh_p$ or $Oh_b$. Since our study involves three fluids we use the subscript $a$ (for air) or $b$ (for bulk water) as prefix to $\mu_r$ and $\rho_r$ to denote the ratio of the parent drop properties and the ambient fluid. For our test conditions, the density ratio, $_ b \rho _ r \approx 0.7 - 0.8$  and $_ a \rho _ r \approx 700 - 800$ are approximately a constant and it may be dropped from further consideration. 
In total, the dimensional analysis presented here shows that the characteristic features of the flow depend only on $Oh_p$ and $_a \mu_r$.  Eqn \ref{eqn:dimvar} now transforms to the form given below, 
\begin{equation} \label{eqn:omegabymu}
\frac{\Omega}{{\Omega}_\mu} \sim f(_a \mu _r, {Oh}_{p})
\end{equation}
As we will see later, eqn \ref{eqn:omegabymu} becomes the basis for obtaining the coalescence regime map. 
\begin{figure*}[htp!]
\begin{center}
  \includegraphics[width=\textwidth]{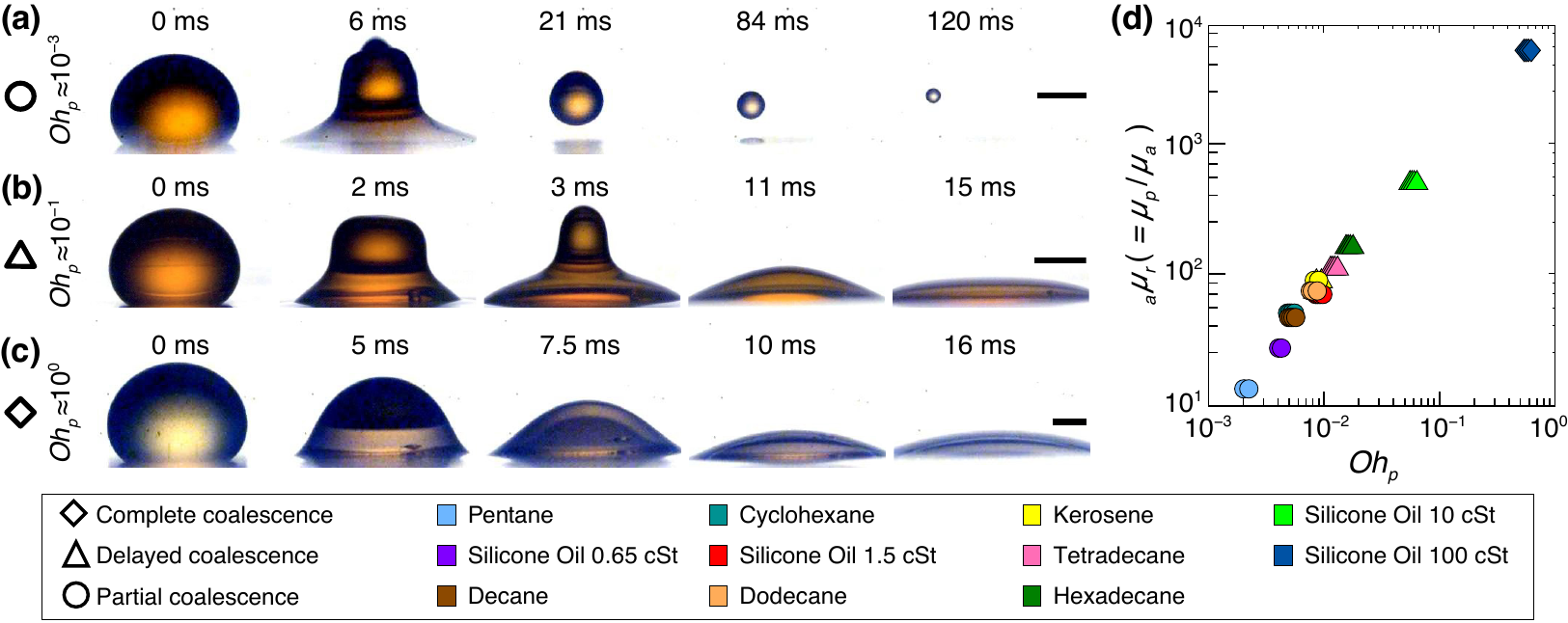}
   \caption{ \label{fig:Figure2} Features of drop coalescence \textbf{(a)} Shows the case of partial coalescence, where the formation of an upward jet leads to the generation of a daughter droplet, Silicone oil (0.65 cSt) ($Oh_p$ = 0.0042). \textbf{(b)} Shows the case of delayed coalescence where formation of a jet is observed in the absence of a daughter droplet Hexadecane ($Oh_p$ = 0.015). \textbf{(c)} shows the case of complete coalescence without jetting. Silicone oil (100 cSt) ($Oh_p$ = 0.60). The existence of a new regime of coalescence termed as \textit{delayed coalescence} is emphasized. (see Section S1, SI Video 1 for videos showing coalescence behavior in the three cases). \textbf{(d)} The regime map delineates the regions of partial, delayed, and complete coalescence as a function of the viscosity ratio ($_a\mu_r = \mu_p/\mu_a$) and the parent drop Ohnesorge number ($Oh_p$). The scale bar in each row corresponds to a length of 1 mm.}
\vspace{-20pt}
\end{center}
\end{figure*}
\subsection{Dominant scales}

Non-dimensionalizing the  drop deformation features would require the dimensional scale $\Omega_{\mu}$ in eqn \ref{eqn:omegabymu} to be determined. However, $\Omega_{\mu}$ is an unknown scale with several different candidates as potential choices. These scales are determined by a balance of the three dominant forces - inertia ($I$), surface tension or capillarity ($C$) and viscosity ($V$) considered two at a time or all three together, each of which results in a different length, velocity and time scale. (Section S4, SI Table S2 shows the different possibilities based on these force balances)

Although choosing $D_p$ (corresponding to the I-C length scale) has been usually preferred in literature \cite{Chen2006a, Aryafar2006a, Blanchette2009a, Kavehpour2015} we explore the applicability of the I-V-C scales first proposed by Eggers and Dupont \cite{Eggers1994}. These scales have recently been identified by Ganan-Calvo \cite{Ganan2017} as being appropriate for analyzing bubble bursting in a viscous liquid and show a remarkable collapse of existing numerical and experimental data \cite{Berny2020, Lai2018, Ganan2017}. Similar to bubble bursting, our problem also involves the the ascent of capillary waves along a phase boundary comprising of air on one side and liquid on the other \cite{Berny2020, Ganan2017}, but in contrast to it they meet at the apex of the drop (anti-parallel to gravity). Further justification on the choice of these scales may be seen in the fact that the intervening air film drains much faster than a liquid film in a two-phase drop coalescence problem \cite{Kavehpour2015}. This implies that we need to seek scales which are much smaller than the ones used earlier thus validating the use of the I-V-C scales. Studies like the one presented by Lai \textit{et al}.\cite{Lai2018} and Berny \textit{et al}.{\cite{Berny2020} may be carried out to additionally bolster our claim. We restrict our attention to showing that these scales work for our experimental data and validate it theoretically using simple scaling arguments. It is of significance that $_a\mu_r$ is constant, if we only consider the ratio of drop and air viscosity and can be dropped out of all scaling relations yielding a scaling dependence of the form, $\Omega/\Omega_\mu \sim Oh_p^n$, where $n$ is any real number.

The scales obtained using the I-V-C scaling can be summarized as given in eqn \ref{eqn:scales} (derived theoretically in Section S5, SI) and will be employed as scales for non-dimensionalising parameters in the following sections.
\begin{equation} \label{eqn:scales}
{l}_{\mu} = \frac{{\mu}_{p}^{2}}{{\rho}_{p}{\gamma}}, \qquad \hspace{0.3em}
{t}_{\mu} = \frac{\mu_p^3}{\rho_p\gamma^{2}}, \qquad \hspace{0.3em}
{v}_{\mu} = \frac{\gamma}{\mu_p}
\end{equation}

Note that often we may use a simple division by length scale $l_{\mu}$ to prove a scaling relation, theoretically even though it precludes the expected explicit inclusion of expression for various forces. This is a consequence of the I-V-C balance which is argued to be applicable to our case given the above considerations. Also, the fact that the capillary waves carve the features on the drop surface and their strength is directly proportional of the liquid properties \cite{Ganan2017} warrants the use of a relatively simple calculation.
\vspace{-10pt}
\subsection{Coalescence Regime Map}\label{ssec:regime_map}
The discussion above has enunciated the significant role of capillary waves in governing drop deformation and coalescence. Analysis of the high-speed imaging videos for the 11 different oils tested (refer Table \ref{tab:Table1}) in this study showed that their coalescence behavior after contacting the bulk water surface could be classified into three main categories as shown in Fig. \ref{fig:Figure2} (a)-(d) (see Section S1, SI Video 01) arranged in increasing order of $Oh_p$ (a measure of increasing viscosity of the drop). 

At very low $Oh_p << 10^{-2}$, the capillary waves are so strong (meaning that they possess higher energy) that they constructively interfere at the apex (topmost point on the deforming drop) and stretch the drop vertically such that a portion of the liquid mass pinches off (daughter droplet) from the parent drop (top row, $Oh_p = 10^{-3}$ Fig. \ref{fig:Figure2}(a)) in what is termed as \textit{partial coalescence}. The process then repeats itself until the drop completely coalesces with the bulk. At much higher  $Oh_p$ ($\gtrapprox 1$), the capillary waves are dampened to the extent that they decay immediately after they appear and complete coalescence (bottom row, $Oh_p = 1$ Fig. \ref{fig:Figure2} (c)) occurs. Interspersed between these two regimes is a new regime of \textit{delayed coalescence} where the capillary waves are strong enough to stretch the drop but relatively weak for a daughter droplet to form (middle row, $Oh_p = 10^{-1}$ Fig. \ref{fig:Figure2}b). To the best of the knowledge of the authors this is the first time that the existence of this regime has been shown to be an intermediate stage between regimes of partial and complete coalescence implying that the transition is not sudden but gradual. Although the role of capillary waves in coalescence of drops with planar surfaces has been well understood \cite{Chen2006b, Kavehpour2015, Blanchette2009a} the existence of \textit{delayed coalescence} has not yet been reported like so in literature despite some allusions \cite{Gilet2007a} to it. Our findings therefore may be succinctly classified into three distinct coalescence regimes $-$ (\textit{i}) Partial coalescence (jetting producing a daughter droplet), (\textit{ii}) Delayed coalescence (jetting without daughter droplet) and, (\textit{iii}) Complete coalescence (no jetting). The terminology of \textit{jetting} refers to the abrupt change in slope of the drop contour from zero to infinity ($t = 3 ms$, frame 3 Fig. \ref{fig:Figure2}(b) and frame 2, Fig. \ref{fig:Figure2}(a)) as against a more gradual change back to zero (see Section S1, SI Video 01). 

In section \ref{sec:danalysis} we showed that the different drop morphologies can be conveniently represented in the terms
of non-dimensional parameters $_a\mu_r$ and $Oh_p$. Upon plotting our experimental data using these variables, regions delimiting the partial, delayed and complete coalescence were clearly observed. A line of best fit drawn including these points gives rise to a linear scaling dependence of the form, $Oh_p \sim \mu_r$. Region 1 (circles) represents the regime of partial coalescence, where both the ${_a\mu}_{r}$ and ${Oh}_{p}$ are relatively low. At higher ${_a\mu}_{r}$ and ${Oh}_{p}$, delayed coalescence is observed, represented by the triangle plots in region 2. Region 3 represents the regime of complete coalescence, where the capillary wave propagation is sufficiently damped. The diamond plots in this region represent drops with relatively high ${_a\mu}_{r}$  and ${Oh}_{p}$. Note that replacing the ambient medium with a different fluid will likely produce lines parallel to the existing data points, although it will not have any effect on the slope of these lines. In the sections that follow we quantify features which are manifestations of these regimes and determine their dependence on liquid (oil) properties.
\vspace{-15pt}
\subsection{Height of drop}
\begin{figure*}
\begin{center}
\includegraphics[width=\textwidth]{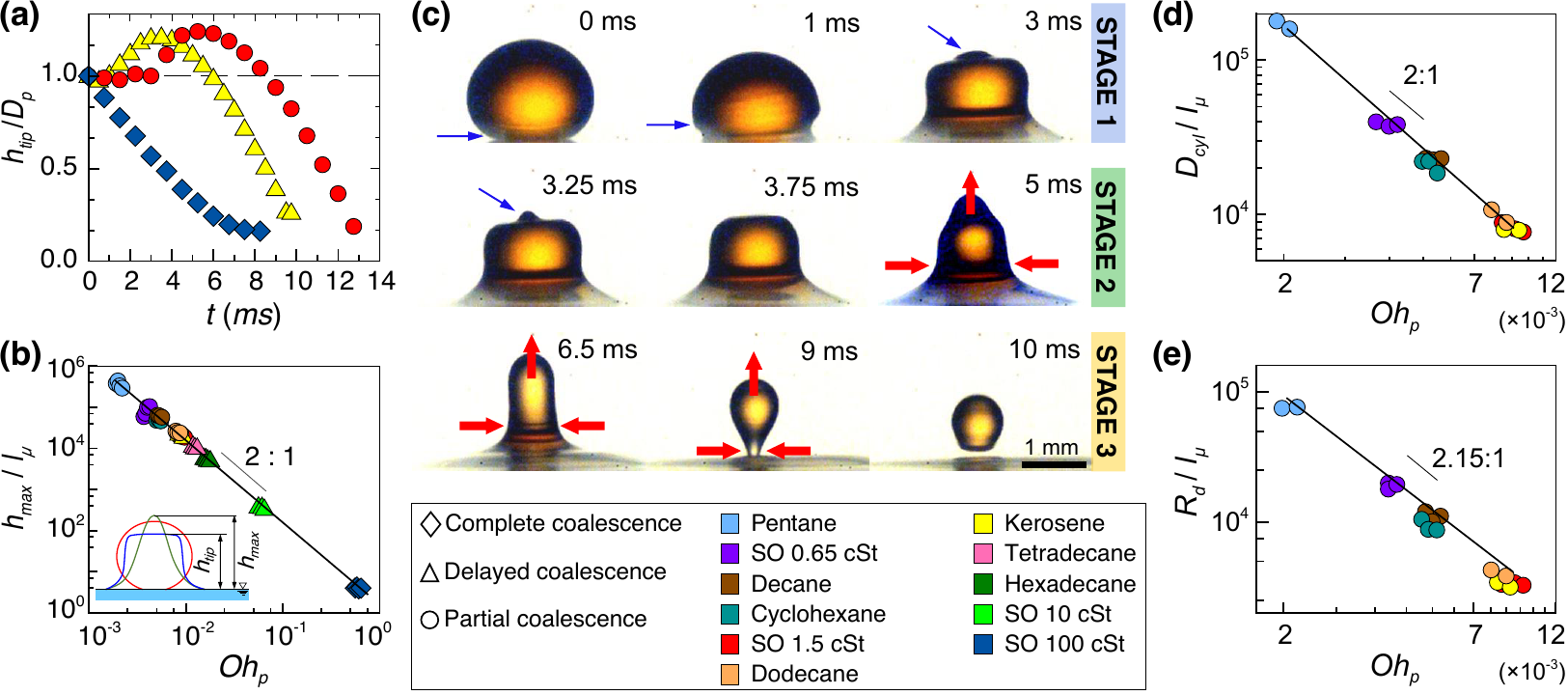}
\caption{\label{fig:Figure3} \textbf{(a)} Variation of $h_{tip}/ D_p$ with time, $t$. A value of greater than 1 means the droplet undergoes \textit{partial} or \textit{delayed} coalescence and exhibits \textit{jetting}. \textbf{(b)} Experimental non$-$ dimensional maximum height ($H_{max}$) of the drop as a function of $Oh_p$ \textbf{(c)} Three stages prior to pinch-off of a daughter droplet in \textit{partial} coalescence: \textbf{Stage 1} - incipience of capillary waves concluding with the crashing of the waves at the apex, \textbf{Stage 2} - tip reversal and gradual horizontal thinning of the drop, \textbf{Stage 3} - Formation of cylindrical liquid entity and its final collapse to produce a daughter droplet. Blue arrows show the movement of the capillary waves along the drop/air interface . Red block arrows indicate the direction of the stretching (vertical) and thinning (horizontal). \textbf{(d)} Experimental scaling for non-dimensional $D_{cyl}$ with $Oh_p$. Noticeably, it is the same as the for non-dimensional  $h_{max}$. \textbf{(e)} Experimental scaling for non-dimensional daughter droplet ($D_d$) with $Oh_p$. Coefficient of determination, $R^2$ for the power law fits is  $\geq$ 0.98 and the scaling exponents reported here have an uncertainty within $\pm$ 5 \%.}
\vspace{-20pt}
\end{center}
\end{figure*}

The foregoing exposition has identified \textit{jetting} or vertical stretching of the drop as one of the differentiating factors between complete and partial/delayed coalescence. Even though visually apparent, an unambiguous quantification in terms of a drop deformation feature to distinguish different regimes is paramount. To achieve this goal, we consider the maximum height of the deformed drop ($h_{max}$, Fig.\ref{fig:Figure1} (b)-(iii), Fig. \ref{fig:Figure2} (a),(b),(c) second column), a natural choice, which also serves as an indicator of the strength of the capillary waves \cite{Gilet2007a}. 

To identify $h_{max}$ for each of the three regimes we plot the variation of $h_{tip}/D_p$ with time, $t$ as shown in Fig. \ref{fig:Figure3}(a). We immediately notice that for complete coalescence the droplet height ($h_{tip}$) never exceeds the droplet diameter (\textit{i}.\textit{e}. $h_{tip}/D_p < 1$) while delayed and partial coalescence are marked by vertical stretching ($h_{tip}/D_p > 1$). This provides the appropriate quantifiable measure to differentiate between partial/delayed and complete coalescence. In order to further distinguish between partial and delayed coalescence we observe that when the rate of vertical stretching exceeds the rate of horizontal thinning, pinch-oﬀ occurs, leading to the formation of a daughter droplet and is symbolic of partial coalescence in contrast to delayed coalescence where there is no daughter droplet formation. The difference in the exact values of $h_{tip}/D_p$ for partial and delayed coalescence is immaterial as the generation of the daughter droplet followed by a cascade of self-similar events is lucid enough to differentiate between the two regimes.

On this note it is pertinent to point out that the horizontal portion in the temporal variation of $h_{tip}/D_p$ (Fig. \ref{fig:Figure3}(a)) for partial coalescence (\tikz\draw[black,fill=red] (0,0) circle (0.65ex);) is an indicator of the period for which there is no change in $h_{tip}$ even though a the air film has ruptured. This is characteristic of partial/delayed coalescence where the nature of the capillary waves is such that for a brief period until they reach the bulk air/drop interface they do not cause any visible change in $h_{tip}$ and increase the total time for coalescence. 

Contrastingly, in the complete coalescence regime $h_{tip}$ decays rapidly thereby reducing this time (discussed at length later in Section \ref{sec:coals_time}) with, $h_{max}$, representing the end of the vertical elongation, hence is a valuable measure and strongly dependent on the drop's viscosity as revealed by our analysis so far. Using the I-V-C length scale ($l_{\mu}$) derived in the previous section it may be expected that the scaled data for $h_{max}$ when expressed in terms of  $Oh_p$  would collapse for all liquids giving rise to a simple power law relation which can then be extended to a wide range of fluids. Fig. \ref{fig:Figure3}(b) precisely shows this dependence where with an increase in ${Oh}_{p}$, the maximum height of the drop decreases and can be mathematically expressed as, 
\begin{equation} \label{eqn:hmax}
\frac{h_{max}}{l_\mu} \sim Oh_p^{-2}
\end{equation}
Based on our observations we see that the maximum height of the drop is of the order of the diameter of the parent drop ($h_{max} \sim D_p$). Non-dimensionalizing ${h}_{max}$ and $D_p$ with the length scale($l_{\mu}$) and recognizing that $\rho_p\gamma R /\mu_p^2 = Oh_p^{-2}$ we obtain the relationship between ${h}_{max}$ and ${Oh}_{p}$ which matches exactly with the experimental values as shown in eqn \ref{eqn:hmax}. 
In the next two sections we delve deeper into mechanism of daughter droplet formation.
\vspace{-10pt}
\subsection{Daughter droplet}

As noted above vertical stretching is critical to the formation of the daughter droplet. This process not only involves extension of the liquid mass upwards but also horizontal thinning at the point where it joins the bulk \cite{Chen2006a, Kavehpour2015, Charles1960a}. Bestriding the free surface boundary are capillary waves shaping the drop contour as they move and elongate the drop. Their journey can be divided into three stages - (Fig. \ref{fig:Figure3}(c)) (\textit{i}) From incipience until they constructively interfere at the apex  (\textit{ii}) tip reversal which ends with a uniform cylindrical element (\textit{iii}) collapse of the cylindrical element (of length, $L_{cyl}$ and diameter, $D_{cyl}$ forming a single daughter droplet. Once a daughter droplet is produced stages (\textit{i}) to (\textit{iii}) are repeated again until the final droplet coalesces with the bulk forming an immiscible liquid film. We observed a cascade of as many as 6 daughter droplets which decreased to 1 at higher $Oh_p$. Cases where one or more droplets formed were very few and not considered in the current analysis.

Against this background it must be noted that droplet ejection from a jet has usually been explained as a consequence of capillary (Rayleigh-Plateau) instability commonly seen in dripping faucets and liquid atomization \cite{Eggers2008}. It is a matter of debate whether the same mechanism governs droplet formation in jets formed due to capillary waves as reported in bursting bubbles \cite{Ganan2017} and partial coalescence \cite{Blanchette2006}. For our case it is evident that the coalescence process is controlled by the competition between the horizontal and vertical rates of collapse and when the rate of horizontal collapse is greater than that of vertical collapse, a daughter droplet is produced. The daughter droplet generation is thus directly dependent on the strength of the capillary waves meeting at the apex of the drop determined by the viscosity of the drop. Fig. \ref{fig:Figure3}(e) shows the scaling dependence of the non-dimensional daughter droplet radius on $Oh_p$ displaying a power law exponent of $-2.5$. 

To derive this relation theoretically we examine the events leading to the pinch-off of the daughter droplet at depicted in stage 3 (Fig. \ref{fig:Figure3}(c)). In this stage a cylindrical mass collapses to form a spherical daughter droplet. The two volumes should be the same to comply with the conservation of mass. Thus the volume of the daughter droplet, $V_d$, ($\sim D_d^3$) is equal to the volume of this cylindrical mass $V_{cyl}$, ($\sim L_{cyl}D_{cyl}^2$) before pinch-off ensues. $L_{cyl}$ is the length of the cylinder before its collapse begins (Fig. \ref{fig:Figure1}) and is equal to $h_{max}$. This implies that the scaling for $L_{cyl}$ and $h_{max}$  should be the same. On similar lines, $D_{cyl}$ scales as $D_{p}$ which amounts to $D_d^3 \sim h_{max}D_{p}^2$. Using the I-V-C length scale we obtain the following scaling relation.
\begin{equation} \label{eqn:dau_drop}
\frac{D_d}{l_\mu} \sim Oh_p^{-2}
\end{equation}

Eqn \ref{eqn:dau_drop} is in agreement with the experimental results(within 7\%)  shown in Fig \ref{fig:Figure3}(d) which yields a scaling relation of $D_d/l_\mu \sim Oh_p^{-2.15}$. Predictably, the radius of the daughter droplet decreases with an increase in $Oh_p$. With increased viscous dissipation, the capillary waves are dampened significantly, resulting in a smaller daughter droplet size. The scaling relation obtained agrees well with previous studies \cite{Kavehpour2015, Aryafar2006a} where it is suggested that the ratio of the radius of the daughter droplet to the radius of the parent drop remains constant at low $Oh_p$ ($\lessapprox 0.1$). This implies that $D_d \sim D_p$ which we show for low $Oh_p$ can be scaled using $\l_{\mu}$ and can also be obtained using our scaling arguments.

\subsection{Coalescence time}\label{sec:coals_time}

When a liquid drop comes in contact with the bulk liquid and starts to spread, it undergoes a change in shape due to the capillary waves straddling its surface, the final outcome of which is the formation of a lens, flat thin film or pseudo-lens (see Section S6, SI for more details). The time interval between the moment the intervening air film ruptures to the formation of the liquid film is defined as the coalescence time, $t_{c}$. Alternatively, in terms of spreading coalescence time ($t_c$) may be envisaged as time that the drop takes to show steady state behavior (refer section \ref{sec:schar}). For this calculation, time $t = 0$ is identified by the moment when the capillary waves are generated and is marked by formation of a prominent crest. To understand clearly the instant of time when coalescence is considered complete we refer to Fig \ref{fig:Figure2} (a), (b), (c) for estimating $t_c$. For partial, delayed and complete coalescence this is approximately 19, 10 and 8 $ms$ respectively (Fig. \ref{fig:Figure2}(a)). In the final coalesced state the drop assumes the shape of a flat liquid film few 10 $\mu m$ in thickness ($S > 0$) or a lens approximately 100 $\mu m$ in thickness at its center ($S < 0$)  (see Section S1, SI Video 01). 

Fig. \ref{fig:Figure4} shows the experimentally measured coalescence time non-dimensionalized using the I-V-C timescale and plotted against $Oh_p$ as given by the scaling relation below (\ref{eqn:c_time}). From Fig. \ref{fig:Figure4}, it is evident that the coalescence time decreases with an increase in $Oh_p$, confirming the dominant role of capillary waves in delaying the vertical collapse, and consequently coalescence of the drop.

To theoretically derive the scaling relation obtained experimentally in Fig. (\ref{fig:Figure4}), the coalescence time ${t}_{c}$ can be considered to scale with the ratio of the maximum height of the drop and the capillary wave velocity $V_{wave}$ which is the velocity with which capillary waves move along the drop/air interface. 
\begin{figure}
\begin{center}
{
\includegraphics[width=\columnwidth]{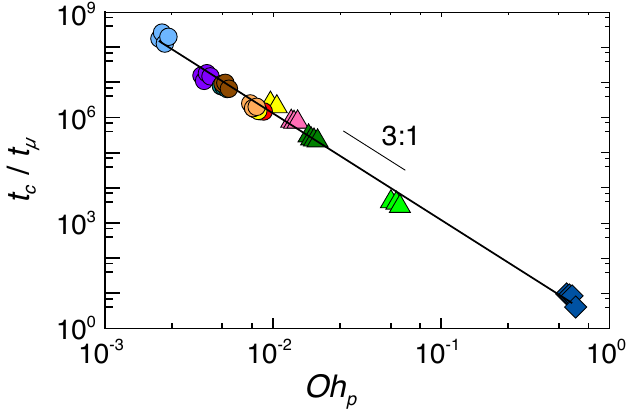}
\caption{\label{fig:Figure4} Variation of coalescence time($t_c$) with $Oh_p$. This is the time required for complete coalescence from initial rupture of air film to the first instant of formation of a lens ($S < 0$) and thin film ($S > 0$). Lower viscosity drops undergoing partial coalescence are seen to take longer to coalesce while high viscosity drops take less time to coalesce. Coefficient of determination, $R^2$ for the power law fit is  $\geq$ 0.98. The scaling exponent of $-$3 has an uncertainty within $\pm$ 5 \%.}
}
\vspace{-22pt}
\end{center}
\end{figure} 
For a viscous fluid/air interface it scales as $\gamma/{\rho_p D_p}$ \cite{Ganan2017} which results in the scaling relation, $V_{wave}/V_{\mu} \sim Oh_p$ (verified experimentally in Section S7, SI). Note that as counter-intuitive as it may seem the scaled velocity of the waves does increase with increasing viscosity. Mathematically, this may be attributed to the scaling, $V_{\mu}$($= \gamma/\mu_p$) which increases with lowering of viscosity. Physically speaking, $V_{wave}/V_{\mu} = \mu_p V_{\mu}/\gamma$ or the wave capillary number, $Ca_{wave}$. It is the relative measure of viscous to capillary forces. A higher value of $Ca_{wave}$ signifies higher dissipation due to viscosity which is the case for higher $Oh_p$. Interestingly, the scaling also succeeds in highlighting the role of viscosity where the raw experimental values (see Section S8, SI  Fig. S3 (b)) are of the same order of magnitude and fail to exhibit a discernible trend. Continuing further we employ the scaling relations for $h_{max}$ and $V_{wave}$ and obtain the following scaling for the non-dimensional coalescence time,
\begin{equation} \label{eqn:c_time}
\frac{t_c}{t_\mu} \sim Oh_p^{-3}
\end{equation}
The coalescence time ($t_c$) has special significance from the standpoint of spread of an oil spill as it marks the boundary of the topological transition from a spherical drop to a planar liquid thin film or a lens. Most of the oil spill mitigation measures have considered only spreading and dynamics when oil is spilled as a film or has taken that form. 

The measurement for $t_c$ presented here can serve to determine this boundary when the spreading kinetics changes dramatically. In the next section we investigate the spreading of the drop in the time($t_c$) that it coalesces with the bulk.
\vspace{-10pt}
\subsection{Spreading characteristics}\label{sec:schar}

In the context of oils spills, the coupling of the coalescence process and the spreading behavior is of immense significance. Most literature has focused on very late time spreading without studying the initial moments of the coalescence process. As such the scaling laws obtained in this late time spreading regime are monotonic and power law \cite{Fay1971, Hoult1972}. The early time spreading behavior for our case  can vary based on its physical properties with profound implications on the final state of the oil slick formed. 

If the drop is very viscous it damps the capillary waves substantially and monotonically spreads to a maximum size and forms a liquid film. If on the other hand, the drop has low viscosity, the inertial force due to the capillary waves and the equivalent interfacial tension force compete with each other, resulting in an oscillatory spreading behavior (see Section S1, SI Video 02). To further understand the behavior of different oils during this transient spreading, we tracked the instantaneous radius of the spreading liquid drop $r\left(t\right)$ at any instant of time is plotted and shown in Fig. \ref{fig:Figure5}(b). 

The montage in Fig. \ref{fig:Figure5}(a) for oils with low $Oh_p$ (= 0.0038) (Hexadecane) is presented as an example to demonstrate the inter-connectivity between coalescence and spreading behavior. A low viscosity (Hexadecane) drop shows an oscillatory behavior, while a drop with higher viscosity (SO 100 cSt) monotonically spreads to a maximum radius. To analyze the most important moments during spreading we looked closely at the features of oscillatory spreading as indicated by the markers at different instants in time in Fig. \ref{fig:Figure5}(b). Note that we only consider the radial growth of the macroscopic film and not the precursor, as may be the case for spreading oils($S>0$). For non-spreading oils($S<0$) such a scenario typically does not arise as there is no microscopic precursor (see Sections S2, S6 SI for more details).

\begin{figure*}
  \includegraphics[width=\textwidth]{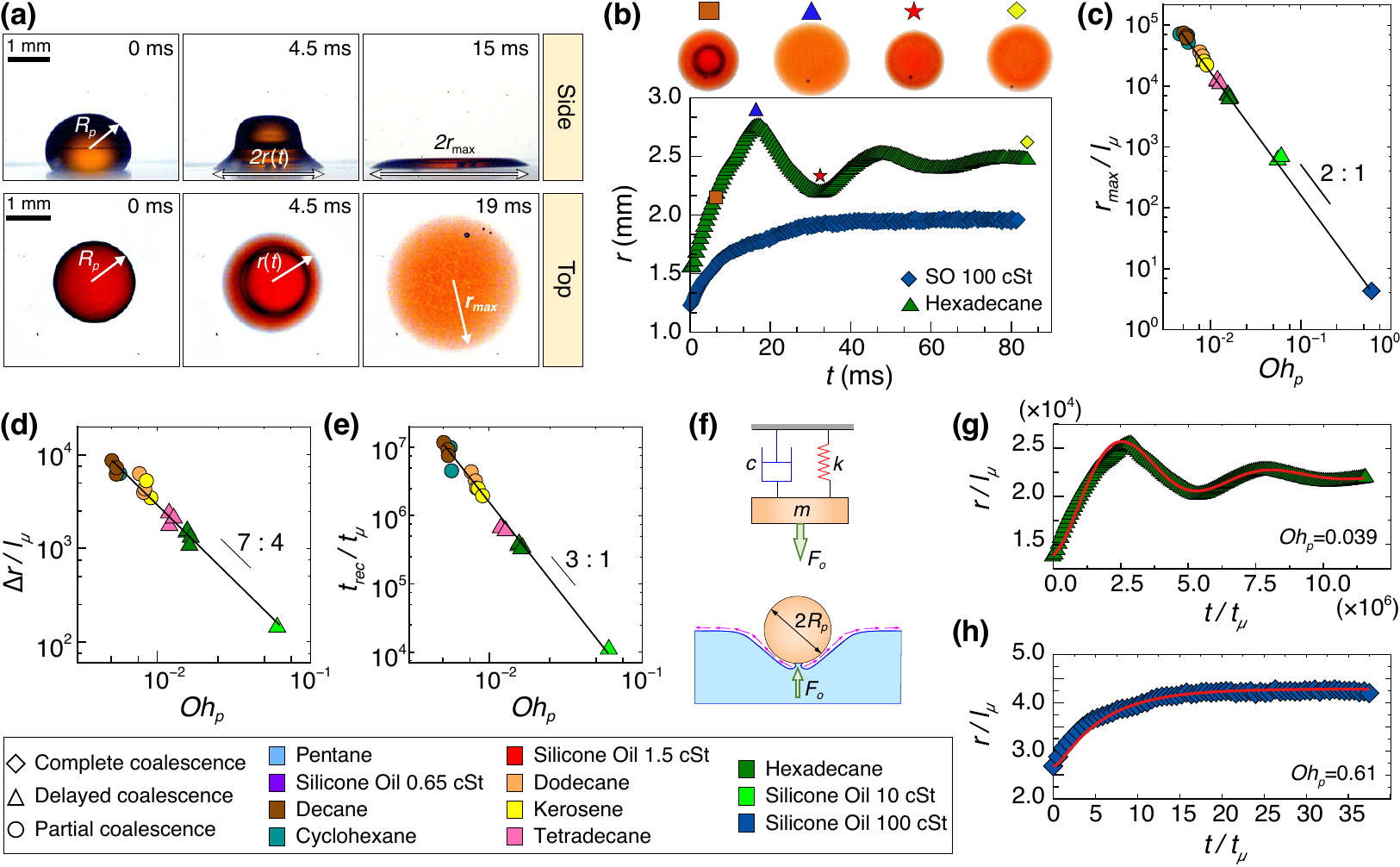}
     \caption{\label{fig:Figure5} Features of drop spreading \textbf{(a)}  Side and top views of the coalescence and spreading for low $Oh_p$ (= 0.038, Hexadecane, \textit{delayed coalescence}). \textbf{(b)} Variation of the spreading radius, $r\left(t\right)$  for oils with low $Oh_p$ (= 0.038, Hexadecane) and relatively high $Oh_p$ (=0.610, Silicone Oil 100 cSt). The symbols indicate different stages in the spreading of a drop. $\begingroup\color{blue}\blacktriangle\endgroup$ - maximum radius $r_{max}$, $\begingroup\color{red}\bigstar\endgroup$ - minimum radius $r_{min}$, $\begingroup\color{olive}\blacklozenge\endgroup$ - steady state radius $r_{ss}$ \textbf{(c)} Experimental scaling of the maximum radius of the film (${r}_{max}/l_\mu \sim Oh_p^{-2}$) and, \textbf{(d)}  Extent of recoil $\Delta r/l_\mu$ (where, $\Delta r = r_{max} - r_{min}$) as a function of the $Oh_p$ ($\Delta r \sim Oh_p^{-7/4}$). Viscosity reduces values of $r_{max}$ and $r_{min}$ and hence their difference. Similarly, \textbf{(e)}  shows the scaling for recoil time $t_{rec}$ as a function of $Oh_p$ ($t_{rec}/l_\mu \sim Oh_p^{-3}$). It is difference between instants of time corresponding to $\begingroup\color{red}\bigstar\endgroup$ and $\begingroup\color{blue}\blacktriangle\endgroup$ \textbf{(f)}  Spring-mass-damper analogue of the spreading drop. The radius of the drop at every time instant $r\left(t\right)$ is measured to determine the features of spreading. The direction of the draining air film is shown by pink arrows. \textbf{(g)}, \textbf{(h)} The solution of the reduced order spring-mass-damper model in comparison with the experimental data of the spreading drop. Red lines denote theoretical predictions while the data points represent experimental measured values. Both \textbf{(g)}  underdamped ($Oh_p$ = 0.026) and \textbf{(h)} overdamped ($Oh_p$ = 0.61) responses of the model match well with the experimental data showing 95$\%$ statistical confidence. Coefficient of determination, $R^2$ for the power law fit is  $\geq$ 0.97. The scaling exponent for power law fits in \textbf{(c)}, \textbf{(d)}  and \textbf{(e)}  have an uncertainty within $\pm$ 8 \%.}
\end{figure*}

\textit{Recoil dynamics}: The first observation one can make from Fig. \ref{fig:Figure5}(a) is that a low viscosity drop spreads to a maximum radius $r_{max}$ ($\begingroup\color{blue}\blacktriangle\endgroup$) due to inertia provided by pinch-off drop, and subsequently recoils to a minimum radius $r_{min}$ ($\begingroup\color{red}\bigstar\endgroup$) before the oscillatory behavior eventually dampens to form a stable oil lens ($\begingroup\color{olive}\blacklozenge\endgroup$) on the surface of water. The difference in radii of these extreme points ($r_{max} - r_{min}$) is the extent of recoil ($\Delta r$, see Sections S2, SI) and the time interval between these two instances is the recoil time ($t_{rec}$). These recoil characteristics are discussed further in this section.

Irrespective of the final spreading behavior (see Sections S2 and S6, SI), the inertial force imparted to the drop as the air film ruptures causes the initial spread to a maximum size. Therefore, the maximum radius attained by the drop $r_{max}$ is an indicator of the inertial force imparted to the drop, ultimately determining the extent of spread of the drop. Opposing this inertial force is the viscous dissipation in the drop due to its inherent viscosity besides the restoring surface tension force which is oscillatory. It is expected that with an increase in the viscous dissipation, the maximum radius of the drop would decrease. Theoretically, the maximum radius ($r_{max}$) attained by the drop is of the order of the diameter of the parent drop ($r_{max} \sim D_p$). Using the I-V-C length scale $l_\mu$ to non-dimensionalize the quantities, we obtain a scaling law for $r_{max}$ in terms of the properties of the drop as $r_{max}/l_\mu \sim Oh_p^{-2}$. This is in close agreement with the experimental data provided in Fig. \ref{fig:Figure5}(c) where the maximum radius attained by the oil lens is non-dimensionalized using $l_\mu$ and plotted against $Oh_p$. 

The restoring force responsible for the retraction of the drop to its minimum radius, $r_{min}$ is provided by the combination of interfacial tensions $\sigma_{oa}$ and $\sigma_{ow}$. Therefore, a high initial driving force propelling the drop to $r_{max}$ does not necessarily result in the larger spread of the drop. During this first cycle of spread and retraction the extent of recoil $\Delta r$ ($= r_{max} - r_{min}$) is indicative of the driving (rupture induced) and restoring interfacial tension force. A higher restoring force results in the reduction of the final lens size, thereby reducing the steady state, late time spread $\left( r_{ss}\right)$. Theoretically, the extent of recoil too, is of the order of the parent drop diameter ($\Delta r \sim D_p$), resulting in the scaling law, $\Delta r/l_\mu \sim Oh_p^{-2}$ obtained by non-dimensionalizing $\Delta r$ with $l_\mu$ and using previously established definition for $Oh_p$. This lies reasonably close to the experimental data in Fig. \ref{fig:Figure5}(d) which displays a dependence on the Ohnesorge number as, $\Delta r/l_\mu \sim Oh_p^{-1.75}$. 

Another associated metric, the recoil time $t_{rec}$ is an important feature as it affects the time it takes for the drop to reach a steady state and hence governing the late time spreading behavior of the drop. This is one of the primary reasons for increased coalescence times for partial coalescence since a large value of $t_{rec}$ implies longer time to achieve coalescence. The scaled recoil time, $t_{rec}$ using the I-V-C time scale is plotted against $Oh_p$ in Fig. \ref{fig:Figure5}(e) displaying the scaling law $t_{rec}/t_\mu \sim Oh_p^{-3}$.

\textit{Initial droplet spreading behavior}: The spreading behavior is highly dependent on the physical properties of the oils. Experimental results in Fig. \ref{fig:Figure5}(a) show that drops with a higher $Oh_p$ display a monotonic increase in the contact radius, $r\left(t\right)$ as they spread while drops with comparatively low $Oh_p$ show a non-monotonic behavior. In order to mathematically model the spreading kinetics we draw an analogy to the forced response of a second order mechanical system. Our motivation being that such behavior has been observed in drops impacting liquid surfaces \cite{Terwagne2013, Feng2016} and therefore, it is not unreasonable to expect such a description to work for our case too. To develop these ideas we consider a mechanical spring-mass-damper system as shown in Fig. \ref{fig:Figure5}(f) with a block of mass \textit{m} $\left(kg \right)$, a spring with spring constant \textit{k} $\left(N/m \right)$, and a dashpot with viscous damping coefficient \textit{c} $\left(N-s/m \right)$. 

To determine the value of these forces as relevant to our case, in the mechanical spring-mass-damper system analogue we examine closely the oil drop behavior as it approaches the bulk liquid. Drainage of the intervening air film establishes contact between the drop and bulk, and a step force of magnitude $F_0$ acts along the oil-air phase boundary after the rupture of the air film. This initial step force ($F_0$) is a consequence of the rupture of the air film and the capillary wave that is generated. The magnitude of this step force is proportional to difference in pressure across the film given by $\left(\sigma_{oa}+\sigma_{aw}\right) D_p$ \cite{Vrij1968}. The dynamic viscosity of the drop resists any deformation and slows down the rate of spread of the drop. The ratio of viscosity of the drop and air is much greater than the ratio of viscosity of the drop and the bulk ($_a\mu_r >> _b\mu_r$) for most of our oils (see Section S2, Table S1, SI), and hence the viscosity of the drop is much more significant than the viscosity of the bulk in damping the spread of the drop. Inertia of the drop aids the spreading while the equivalent interfacial tension $\gamma$ provides the restoring force. We combine these various elements to write a second order linear differential eqn for the contact radius of the drop at any instant of time, $r\left(t\right)$ as,
\begin{equation} \label{eqn:EOM_drop}
m\frac{{d}^{2}r}{dt^2} + c\frac{dr}{dt} + kr = F_0 
\end{equation}
Where, the initial radius $r(0) = R_p$ equal to the radius of the parent drop, and initial velocity $v(0) = 0$ are the initial conditions required to solve for $r\left(t\right)$. In eqn \ref{eqn:EOM_drop}, the mass of the drop $m$ is given as $m = \rho_p \pi D_p^3/6$, the coefficient of viscous damping force $c$ is written as $\mu_pD_p$, and the spring coefficient due to equivalent interfacial tension $k$ can be expressed as $k = \gamma D_p$.

As the inertial, viscous, and surface tension forces are all equally dominant, we consider the appropriate I-V-C length and time scales to transform the equation of motion (eqn \ref{eqn:EOM_drop}) into its non-dimensional form. For the sake of brevity, the  dimensionless variables for time $\widetilde{t}$ and instantaneous radius $\widetilde{r}$ are defined as $\widetilde{t} = t/t_\mu$ and $\widetilde{r}(t) = r(t)/l_\mu$. Recognizing the previously defined quantities $m$, $c$, and $k$, we write the non-dimensional form of eqn \ref{eqn:EOM_drop} for the spreading of the drop as given below.

\begin{equation} \label{eqn:EOM_nondim}
{Oh^{-6}_p} \frac{d^2\widetilde{r}}{d\widetilde{t}^2} + {Oh^{-2}_p}\frac{d\widetilde{r}}{d\widetilde{t}} + \widetilde{r} = \widetilde{F}
\end{equation}

Here, $Oh_p = \mu_p/\sqrt{\rho_p\gamma D_p}$ is the Ohnesorge number of the parent drop and $\widetilde{F} = F_0/\gamma D_p$ is the dimensionless form of the force obtained by normalizing the initial step force $F_0$ with the equivalent interfacial tension force $\gamma D_p$ acting on the drop at the time of contact.

The solution of the non-homogeneous ordinary differential equation (ODE) (eqn \ref{eqn:EOM_nondim}) is given by $\widetilde{r}(t) = A_1 e^{s_1 t} + A_2 e^{s_2 t} + \widetilde{F}$, where $s_1$ and $s_2$ are the roots of the characteristic equation of the system, and $A_1$ and $A_2$ are constants obtained from the initial conditions for displacement and velocity (See Section S9, SI). The characteristic equation of the system is written as $s^2 + 2 \zeta \omega_n s + {\omega}_n^2 = 0$, where $\omega_n$ is the undamped natural frequency given by $\omega_n = \sqrt{k/m}$, and $\zeta$ is the viscous damping ratio given by $\zeta = c/c_{cr}$. $c_{cr}$ is the critical damping value given as $c_{cr} = 2\sqrt{km}$.

The roots of the polynomial, given by $s_{1,2} = -\zeta \omega_n \mp \omega_n\sqrt{\zeta^2 - 1}$, are highly dependent on the damping ratio $\zeta$ and so we distinguish between the roots for an underdamped system with $0 < \zeta < 1$ and an overdamped system with $\zeta > 1$. For an underdamped system, the roots are $s_{1,2} =  -\zeta \omega_n \mp i\omega_n\sqrt{1 - \zeta^2}$ where $i = \sqrt{-1}$, while for an overdamped system, the roots are given by $s_{1,2} = \omega_n(-\zeta \pm \sqrt{\zeta^2 - 1})$. Thus, the solutions for the underdamped and overdamped systems are given by the expressions, \ref{eqn:solution_underdamped} and \ref{eqn:solution_overdamped} respectively. 
\begin{equation} \label{eqn:solution_underdamped}
\widetilde{r}(t) = e^{-\zeta \omega_n t} (C_{u1}cos(\omega_d t) + C_{u2}sin(\omega_d t)) + \widetilde{r}_{ss}
\end{equation}
\begin{equation} \label{eqn:solution_overdamped}
\widetilde{r}(t) = e^{-\zeta \omega_n t} (C_{o1}cosh(\omega_\ast t) + C_{o2}sinh(\omega_\ast t)) + \widetilde{r}_{ss}
\end{equation}
where $\widetilde{r}_{ss}$ represents the non-dimensionalized steady state radius of the film after it has stabilized (see figure \ref{fig:Figure5}). The constants $C_{u1}, \;C_{u2},\;C_{o1}$ and $C_{o2}$ are obtained from the initial conditions (at $t=0$, $\frac{d\widetilde{r}}{dt}\left(0\right)$ = $v_0 = 0$, $r\left(0\right) = R_p$) and are given by, $C_{u1} = C_{o1} = \frac{R_p - r_{ss}}{l_\mu}$, $C_{u2} = \frac{v_o + \zeta \omega_n C_{u1}}{\omega_{ud}\;l_\mu}$ and $C_{o2} = \frac{v_o + \zeta \omega_n C_{o1}}{\omega_{od}\;l_\mu}$. 

Further, the natural frequency and the damping coefficient can be written in terms of $Oh_p$ as $\omega_n = Oh_p^6$ and $\zeta = Oh_p/2$. The terms $\omega_{ud}$ and $\omega_{od}$ represent the damped frequency and overdamping coefficient for the underdamped and overdamped systems respectively. They are defined as $\omega_{ud} = \omega_n\sqrt{1-\zeta^2}$ and $\omega_{od} = \omega_n\sqrt{\zeta^2 - 1}$. The solution for eqn \ref{eqn:EOM_nondim} is found for the underdamped and overdamped cases using the $Oh_p$ of the drop and the initial conditions (see Section S9, SI) and plotted alongside the experimental data in Fig. \ref{fig:Figure5}(g),(h). $\mathcal{O}$($1$) prefactors ranging between 3.46 and 6.87 are used in the estimation of the damping frequency ($\omega_d$). The theoretical results match the experimental data closely with $95\%$ statistical confidence. The spreading of the drops can thus be predicted accurately using the response of a second order mechanical system.

\section{Discussion and Conclusions}
To summarize, in this work, we show that the coalescence behavior of an oil drop in a three-phase system depends highly on the physical properties of the drop and can be characterized by the Ohnesorge number of the parent drop, $Oh_p$, namely (\textit{i}) Partial  ($Oh_p \approx 10^{-3}$), (\textit{ii}) Delayed ($Oh_p \approx 10^{-2}$), and (\textit{iii}) Complete coalescence ($Oh_p \approx 10^{-1}$). We also find that initial time spreading follows an oscillatory behavior before reaching its final shape. An analytical model based on the forced response of a second order mechanical system is shown to be predict the spreading behavior of the drop well.

For the first time, the effect of coalescence on spreading is investigated using both experiments and theory. To show that the experimental results may suitably be extended to a wide range of oils, a novel application of the I-V-C scale is pursued. Through these implementations the significant role of the drop viscosity and initial spreading coefficient on the coalescence and spreading behavior is hypothesized.

While delayed coalescence has been observed for two-phase oil-water systems before \cite{Gilet2007a}, here we have shown, its existence as an intermediate regime between partial and complete coalescence in a three-phase system and unreported heretofore. In contrast to previous works \cite{Aryafar2006a, Kavehpour2015}, mainly involving two-phases we have also shown that the dimensional scales obtained via an inertial, viscous, and capillary force balance (I-V-C) are the most suitable scales for analyzing different features generated by the moving capillary waves such as maximum drop height, apex velocity, and coalescence time, leading up to the generation of the daughter droplet for this kind of three-phase drop-interface system. This is a significant advancement from previous works \cite{Chen2006a, Chen2006b, Aryafar2006a} which have not analyzed spreading accompanied by coalescence and are restricted to two-phase systems. Furthermore, our results show that the initial spreading of the drop differs markedly from the late time behavior \cite{Hoult1972, Huh1975}, and is influenced by $Oh_p$. 

Our manuscript provides the foundation for further studies on oil-spreading dynamics. For example, future work may focus on studying the coalescence and spreading behavior in presence of surfactants/nanoparticles in oil/water/both media. The difference in monotonic versus non-monotonic spreading of different oils could inform development of oil-water separation techniques. Consequently, we anticipate that the results of this work can guide studies on oil-water interactions specifically aiding determination of control parameters and optimization for applications such as targeted drug delivery, production of emulsions, enhanced oil recovery, dispersion of oil during a spill and measures to mitigate its spread.
\vspace{-11pt}
\bibliographystyle{apsrev4-2}
\bibliography{Main_References}

\end{document}